\documentclass[12pt]{article}
\usepackage{epsfig}
\def\be{\begin{equation}}
\def\ee{\end{equation}}
\def\bea{\begin{eqnarray}}
\def\eea{\end{eqnarray}}
\usepackage{graphicx}

\catcode`\@=11
\def\lsim{\mathrel{\mathpalette\@versim<}}
\def\gsim{\mathrel{\mathpalette\@versim>}}
\def\@versim#1#2{\vcenter{\offinterlineskip
\ialign{$\m@th#1\hfil##\hfil$\crcr#2\crcr\sim\crcr } }}
\catcode`\@=12
\usepackage{axodraw}

\catcode`@=12
\begin{document}
\thispagestyle{empty}
\begin{flushright}
UCRHEP-T461\\
February 2009\
\end{flushright}
\vspace{0.3in}
\begin{center}
{\LARGE \bf Generalized Lepton Number and\\
Dark Left-Right Gauge Model\\}
\vspace{0.8in}
{\bf Shaaban Khalil$^{a,b}$, Hye-Sung Lee$^c$, and Ernest Ma$^c$\\}
\vspace{0.2in}
{\sl $^a$ Centre for Theoretical Physics, The British University in Egypt,\\
El Sherouk City, Postal No.~11837, P.O.~Box 43, Egypt\\}
\vspace{0.1in}
{\sl $^b$ Department of Mathematics, Ain Shams University,\\
Faculty of Science, Cairo 11566, Egypt\\}
\vspace{0.1in}
{\sl $^c$ Department of Physics and Astronomy, University of California,\\
Riverside, California 92521, USA\\}
\end{center}
\vspace{0.8in}
\begin{abstract}\
In a left-right gauge model of particle interactions, the left-handed fermion
doublet $(\nu,e)_L$ is connected to its right-handed counterpart $(n,e)_R$
through a scalar bidoublet so that $e_L$ pairs with $e_R$, and $\nu_L$ with
$n_R$ to form mass terms.  Suppose the latter link is severed without
affecting the former, then $n_R$ is not the mass partner of $\nu_L$, and as
we show in this paper, becomes a candidate for dark matter which is relevant
for the recent PAMELA and ATIC observations.  We accomplish this in a specific
nonsupersymmetric model, where a generalized lepton number can be defined, so
that $n_R$ and $W^\pm_R$ are odd under $R \equiv (-1)^{3B+L+2j}$.  Fermionic
leptoquarks are also predicted.

\end{abstract}

\newpage
\baselineskip 24pt

\noindent \underline{\it Introduction}~:~ It was recognized 22 years ago
\cite{m87,bhm87} that an unconventional left-right gauge extension of the
standard model (SM) of particle interactions is possible, with a number of
desirable properties. This has become known in the literature as the
alternative left-right model (ALRM) \cite{hr89}.  It differs from the
conventional left-right model (LRM) \cite{dgko91} in that tree-level
flavor-changing neutral currents are naturally absent so that the $SU(2)_R$
breaking scale may be easily below a TeV, allowing both the charged $W_R^\pm$
and the extra neutral $Z'$ gauge bosons to be observable at the large hadron
collider (LHC).  In this paper, we propose a new variant of this extension
which we call the dark left-right model (DLRM).  It predicts the
{\it parallel} existence of neutrinos and {\it scotinos}, i.e. fermionic
dark-matter candidates, as explained below.

\noindent \underline{\it Model}~:~ Consider the gauge group $SU(3)_C \times
SU(2)_L \times SU(2)_R \times U(1)$.  The conventional leptonic assignments
are $\psi_L = (\nu,e)_L \sim (1,2,1,-1/2)$ and  $\psi_R = (\nu,e)_R \sim
(1,1,2,-1/2)$.  Hence $\nu$ and $e$ obtain Dirac masses through the
Yukawa terms $\overline{\psi}_L \Phi \psi_R$ and $\overline{\psi}_L
\tilde{\Phi} \psi_R$, where $\Phi = (\phi_1^0, \phi_1^-; \phi_2^+,
\phi_2^0) \sim (1,2,2,0)$ is a Higgs bidoublet and $\tilde{\Phi} =
\sigma_2 \Phi^* \sigma_2 = (\overline{\phi_2^0},-\phi_2^-;-\phi_1^+,
\overline{\phi_1^0})$ transforms in the same way.  Both $\langle \phi_1^0
\rangle$ and $\langle \phi_2^0 \rangle$ contribute to $m_\nu$ and $m_e$,
and similarly $m_u$ and $m_d$ in the quark sector, resulting thus in the
appearance of tree-level flavor-changing neutral currents \cite{gw77}.

Suppose the term $\overline{\psi}_L \tilde{\Phi} \psi_R$ is forbidden by a
symmetry, then the same symmetry may be used to maintain $\langle \phi_1^0
\rangle = 0$ and only $e$ gets a mass through $\langle \phi_2^0 \rangle
\neq 0$.  At the same time, $\nu_L$ and $\nu_R$ are not Dirac mass partners,
and since they are neutral, they can in fact be completely different
particles with independent masses of their own.  Whereas $\nu_L$ is clearly
the neutrino we observe in the usual weak interactions, $\nu_R$ can now be
something else entirely.  Here we rename $\nu_R$ as $n_R$ and show that
it may in fact be a {\it scotino}, i.e. a fermionic dark-matter candidate.

We impose a new global U(1) symmetry $S$ in such a way that the spontaneous
breaking of $SU(2)_R \times S$ will leave the combination $L = S - T_{3R}$
unbroken.  We then show that $L$ is a generalized lepton number, with
$L=1$ for the known leptons, and $L=0$ for all known particles which
are not leptons.  Our model is nonsupersymmetric, but it may be rendered
supersymmetric by the usual procedure which takes the SM to the MSSM
(minimal supersymmetric standard model).  Under $SU(3)_C \times SU(2)_L
\times SU(2)_R \times U(1) \times S$, the fermions transform as shown
in Table 1. Note the necessary appearance of the exotic quark $h$, which
will turn out to carry lepton number as well.

\begin{table}[htb]
\caption{Fermion content of proposed model.}
\begin{center}
\begin{tabular}{|c|c|c|}
\hline
Fermion & $SU(3)_C \times SU(2)_L \times SU(2)_R \times U(1)$ & $S$ \\
\hline
$\psi_L = (\nu,e)_L$ & $(1,2,1,-1/2)$ & $1$ \\
$\psi_R = (n,e)_R$ & $(1,1,2,-1/2)$ & $1/2$ \\
\hline
$Q_L = (u,d)_L$ & $(3,2,1,1/6)$ & $0$ \\
$Q_R = (u,h)_R$ & $(3,1,2,1/6)$ & $1/2$ \\
$d_R$ & $(3,1,1,-1/3)$ & $0$ \\
$h_L$ & $(3,1,1,-1/3)$ & $1$ \\
\hline
\end{tabular}
\end{center}
\end{table}

The scalar sector consists of one bidoublet and two doublets:
\begin{equation}
\Phi = \pmatrix{\phi_1^0 & \phi_2^+ \cr \phi_1^- & \phi_2^0}, ~~~
\Phi_L = \pmatrix{\phi_L^+ \cr \phi_L^0}, ~~~ \Phi_R = \pmatrix{\phi_R^+
\cr \phi_R^0},
\end{equation}
as well as two triplets for making $\nu$ and $n$ massive separately:
\begin{equation}
\Delta_L = \pmatrix{\Delta_L^+/\sqrt{2} & \Delta_L^{++} \cr \Delta_L^0 &
-\Delta_L^+/\sqrt{2}}, ~~~ \Delta_R = \pmatrix{\Delta_R^+/\sqrt{2} &
\Delta_R^{++} \cr \Delta_R^0 & -\Delta_R^+/\sqrt{2}}.
\end{equation}
Their assignments under $S$ are listed in Table 2.

\begin{table}[htb]
\caption{Scalar content of proposed model.}
\begin{center}
\begin{tabular}{|c|c|c|}
\hline
Scalar & $SU(3)_C \times SU(2)_L \times SU(2)_R \times U(1)$ & $S$ \\
\hline
$\Phi$ & $(1,2,2,0)$ & $1/2$ \\
$\tilde{\Phi} = \sigma_2 \Phi^* \sigma_2$ & $(1,2,2,0)$ & $-1/2$ \\
$\Phi_L$ & $(1,2,1,1/2)$ & $0$ \\
$\Phi_R$ & $(1,1,2,1/2)$ & $-1/2$ \\
$\Delta_L$ & $(1,3,1,1)$ & $-2$ \\
$\Delta_R$ & $(1,1,3,1)$ & $-1$ \\
\hline
\end{tabular}
\end{center}
\end{table}

The Yukawa terms allowed by $S$ are then $\overline{\psi}_L \Phi \psi_R$,
$\overline{Q}_L \tilde{\Phi} Q_R$, $\overline{Q}_L \Phi_L d_R$, $\overline{Q}_R
\Phi_R h_L$, $\psi_L \psi_L \Delta_L$, and $\psi_R \psi_R \Delta_R$, whereas
$\overline{\psi}_L \tilde{\Phi} \psi_R$, $\overline{Q}_L \Phi Q_R$, and
$\overline{h}_L d_R$ are forbidden.  Hence $m_e$, $m_u$ come from $v_2 =
\langle \phi_2^0 \rangle$, $m_d$ comes from $v_3 = \langle \phi_L^0 \rangle$,
$m_h$ comes from $v_4 = \langle \phi_R^0 \rangle$, $m_\nu$ comes from $v_5 =
\langle \Delta_L^0 \rangle$, and $m_n$ comes from $v_6 = \langle \Delta_R^0
\rangle$.  This structure shows clearly that flavor-changing neutral currents
are guaranteed to be absent at tree level.

\noindent \underline{\it Higgs structure}~:~ We now show that $v_1 = \langle
\phi_1^0 \rangle = 0$ is a solution of the Higgs potential which leaves the
combination $L = S - T_{3R}$ unbroken, even as $SU(2)_L \times SU(2)_R \times
U(1) \times S$ is broken all the way down to $U(1)_{em}$.  The generalized
lepton number $L$ remains 1 for $\nu$ and $e$, and 0 for $u$ and $d$, but
the new particle $n$ has $L=0$ and $h$ has $L=1$, whereas $W_R^\pm$ has
$L=\mp 1$ and $Z'$ has $L=0$, etc.  As neutrinos acquire Majorana masses,
$L$ is broken to $(-)^L$.  The generalized $R$ parity is then defined in
the usual way, i.e. $(-)^{3B+L+2j}$.  The known quarks and leptons have even
$R$, but $n$, $h$, $W_R^\pm$, and $\Delta_R^\pm$ have odd $R$.  Hence the
lightest $n$ can be a viable dark-matter candidate if it is also the lightest
among all the particles having odd $R$.  Note that $R$ parity has now been
implemented in a {\it nonsupersymmetric} model.

The Higgs potential of $\Phi$, $\Phi_L$, $\Phi_R$, $\Delta_L$, and $\Delta_R$
consists of many terms.  Considered as a function of their vacuum expectation
values, its minimum is of the form
\begin{equation}
V = \sum_i m_i^2 v_i^2 + {1 \over 2} \sum_{i,j} \lambda_{ij} v_i^2 v_j^2 +
2 \mu_R v_4^2 v_6 + 2 \mu_2 v_2 v_3 v_4 + 2 \lambda' v_1^2 v_5 v_6.
\end{equation}
The last three terms of $V$ come from the allowed terms ${\Phi}_R^\dagger
\Delta_R \tilde{\Phi}_R$, $\Phi_L^\dagger \Phi \Phi_R$, and 
$Tr(\tilde{\Phi}^\dagger
\Delta_L \Phi \Delta_R^\dagger)$, whereas the terms $Tr(\Phi
\tilde{\Phi}^\dagger)$ $(v_1 v_2)$, $\Phi_L^\dagger \tilde{\Phi}
\Phi_R$ ($v_1 v_3 v_4$), $\Phi_L^\dagger \Phi \tilde{\Phi}^\dagger
\Phi_L$ ($v_1 v_2 v_3^2$), $\Phi_R^\dagger \Phi^\dagger \tilde{\Phi}
\Phi_R$ ($v_1 v_2 v_4^2$), $Tr(\Phi^\dagger \Delta_L \tilde{\Phi}
\Delta_R^\dagger)$ ($v_2^2 v_5 v_6$), and $Tr(\Phi^\dagger \Delta_L \Phi
\Delta_R^\dagger)$ ($v_1 v_2 v_5 v_6$) are all forbidden. From the conditions
$\partial V/\partial v_i = 0$,  it is clear that a solution exists for which
$v_1=0$ because $V$ is a function of $v_1^2$ only, even if all other $v_i$'s
are nonzero, provided of course that $\partial^2 V/\partial v_1^2 > 0$
which is satisfied for a range of values in parameter space.  It also shows
that the only residual global symmetry of $V$ is $L$.

The breaking of $SU(2)_R \times U(1) \times S$ to $U(1)_Y \times L$ is
accomplished by $v_4 \neq 0$ and $v_6 \neq 0$.  Note here that $\phi_R^0$ has
$L = -1/2 - (-1/2) = 0$ and $\Delta_R^0$ has $L = -1 - (-1) = 0$. The
subsequent breaking of $SU(2)_L \times U(1)_Y$ to $U(1)_{em}$ occurs through
$v_2 \neq 0$ and $v_3 \neq 0$.  Note here that $\phi_2^0$ has $L = 1/2 - 1/2
= 0$ and $\phi_L^0$ has $L = 0 - 0 = 0$.  At this stage, neutrinos are
massless because $v_5 = 0$ is protected by $L$.  We now add the
dimension-three soft term $\mu_L \tilde{\Phi}_L^\dagger \Delta_L \Phi_L$ which
breaks $L$ explicitly by two units, so that a small
$v_5 \simeq -\mu_L v_3^2/m_5^2$ is induced \cite{ms98} and neutrinos acquire
mass.  Here $\mu_L$ may be natuarlly small because it breaks $L$ to $(-)^L$.
Note that $n$ becomes massive in an exactly parallel way through
$v_6$ but without breaking lepton number.  Note also that the observed 
baryon asymmetry of the Universe is obtainable through leptogenesis from 
$\Delta_L$ decay \cite{ms98}.

\noindent \underline{\it Gauge sector}~:~ Since $e$ has $L=1$ and $n$ has
$L=0$, the $W_R^+$ of this model must have $L = S - T_{3R} = 0 - 1 = -1$.
This also means that unlike the conventional LRM, $W_R^\pm$ does not mix
with the $W_L^\pm$ of the SM at all.  This important property allows the 
$SU(2)_R$ breaking scale to be much lower than it would be otherwise, 
as explained already 22 years ago \cite{m87,bhm87}.  Assuming that 
$g_L = g_R$ and let $x \equiv \sin^2 \theta_W$, then the neutral gauge 
bosons of the DLRM (as well as the ALRM) are given by
\begin{equation}
\pmatrix{A \cr Z \cr Z'} = \pmatrix{\sqrt{x} & \sqrt{x} & \sqrt{1-2x} \cr
\sqrt{1-x} & -x/\sqrt{1-x} & -\sqrt{x(1-2x)/(1-x)} \cr 0 &
\sqrt{(1-2x)/(1-x)} & -\sqrt{x/(1-x)}} \pmatrix{W_L^0 \cr W_R^0 \cr B}.
\end{equation}
Whereas $Z$ couples to the current $J_{3L} - x J_{em}$ with coupling
$e/\sqrt{x(1-x)}$ as in the SM, $Z'$ couples to the current
\begin{equation}
J_{Z'} = x J_{3L} + (1-x) J_{3R} - x J_{em}
\end{equation}
with the coupling $e/\sqrt{x(1-x)(1-2x)}$.  The masses of the gauge bosons
are given by
\begin{eqnarray}
&& M_{W_L}^2 = {e^2 \over 2x} (v_2^2 + v_3^2), ~~~ M_Z^2 = {M_{W_L}^2 \over 1-x},
~~~ M_{W_R}^2 = {e^2 \over 2x} (v_2^2 + v_4^2 + 2v_6^2), \\
&& M_{Z'}^2 = {e^2 (1-x) \over 2x  (1-2x)} (v_2^2 + v_4^2 + 4v_6^2) -
{x^2 M_{W_L}^2 \over (1-x)(1-2x)},
\end{eqnarray}
where zero $Z-Z'$ mixing has been assumed for simplicity, using the condition
\cite{bhm87} $v_2^2/(v_2^2+v_3^2) = x/(1-x)$.  Note that in the ALRM,
$\Delta_R$ is absent, hence $v_6=0$ in the above.  Also, the assignment
of $(\nu,e)_L$ there is different, hence the $Z'$ of the DLRM is not identical
to that of the ALRM.  At the LHC, if a new $Z'$ exists which couples to
both quarks and leptons, it will be discovered with relative ease.  Once
$M_{Z'}$ is determined, then the DLRM predicts the existence of $W_R^\pm$
with a mass in the range
\begin{equation}
{(1-2x) \over 2(1-x)} M_{Z'}^2 + {x \over 2(1-x)^2} M_{W_L}^2 < M_{W_R}^2 <
{(1-2x) \over (1-x)} M_{Z'}^2 + {x^2 \over (1-x)^2} M_{W_L}^2.
\end{equation}
In the ALRM, since $v_6=0$, $M_{W_R}$ takes the value of the upper limit
of this range.  The prediction of $W_R^\pm$ in addition to $Z'$ distinguishes
these two models from the multitude of other proposals with an extra $U(1)'$
gauge symmetry.

\noindent \underline{\it Bounds on $SU(2)_R$}~:~   Using Eq.~(5) and assuming
$\Gamma_{Z'} = 0.05~M_{Z'}$, which is about twice what it would be if $Z'$
decays only into SM fermions, we compute the cross section
$\sigma ( p \bar{p} \to Z' \to e^+ e^-)$ at a center-of-mass energy
$E_{cm} = 1.96$ TeV.  We show this in Fig.~1 (left) together with the
lower bound on $M_{Z'}$ from the Tevatron search, based on an integrated
luminosity of $L = 2.5$ fb$^{-1}$ \cite{cdf}.  We obtain thus $M_{Z'} > 850$
GeV, and using Eq.~(8), $M_{W_R} > 500$ GeV.  In Fig.~1 (right) we show the
discovery reach of the LHC ($E_{cm} = 14$ TeV) for observing 10 such
events with the cuts $p_T > 20$ GeV and $|\eta| < 2.4$ for each lepton, and
$|m_{e^- e^+} - M_{Z'}| < 3\Gamma_{Z'}$.  The dominant SM background from
$\gamma/Z$ (Drell-Yan) is negligible in this case.  With an integrated
luminosity of $L = 1$ fb$^{-1}$ (10 fb$^{-1}$), up to $M_{Z'} \sim 1.5$
TeV (2.4 TeV) may be probed.

\begin{figure}[htb]
\begin{center}
\includegraphics[width=0.45\textwidth]{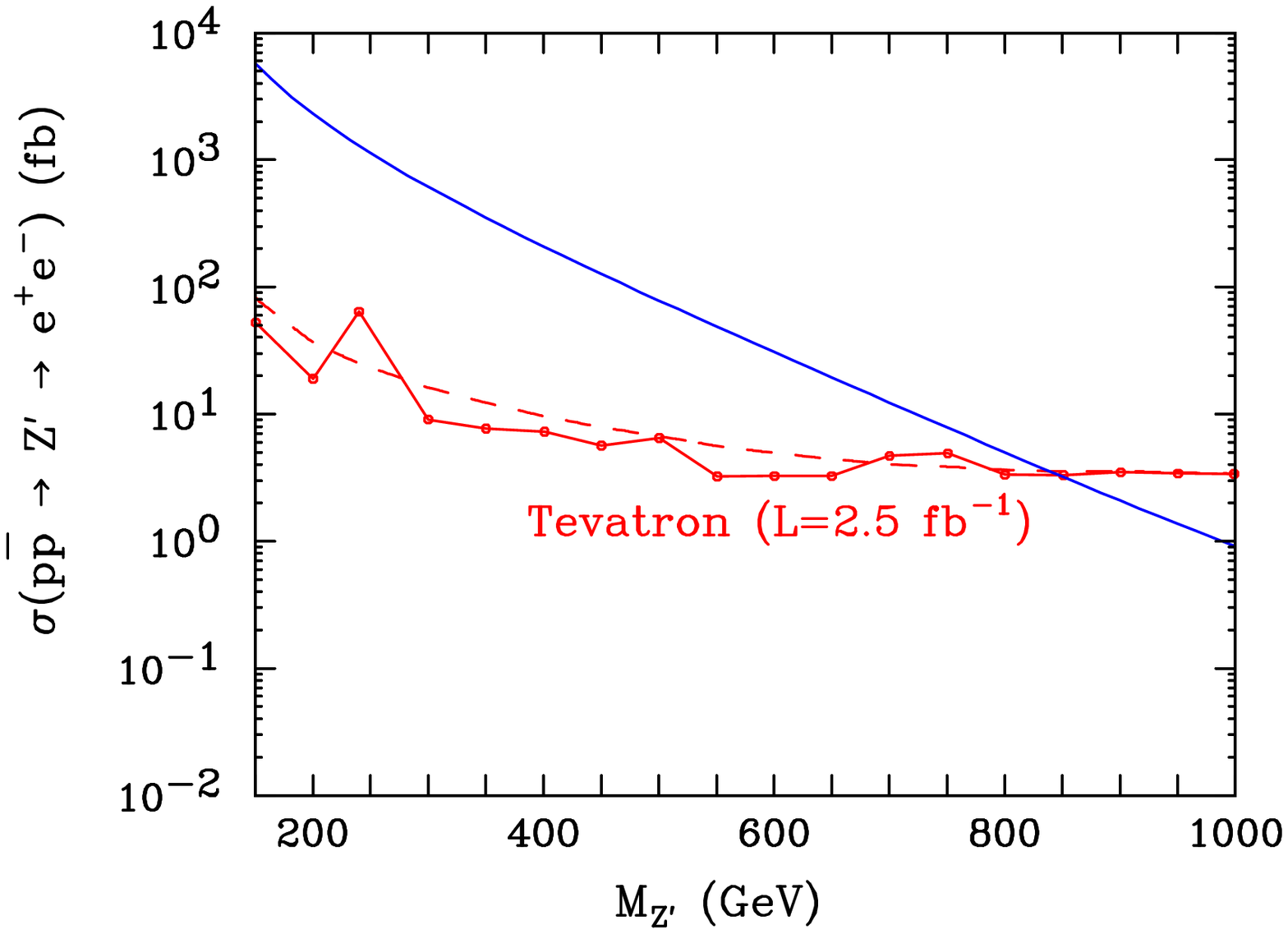} ~~~~~~
\includegraphics[width=0.45\textwidth]{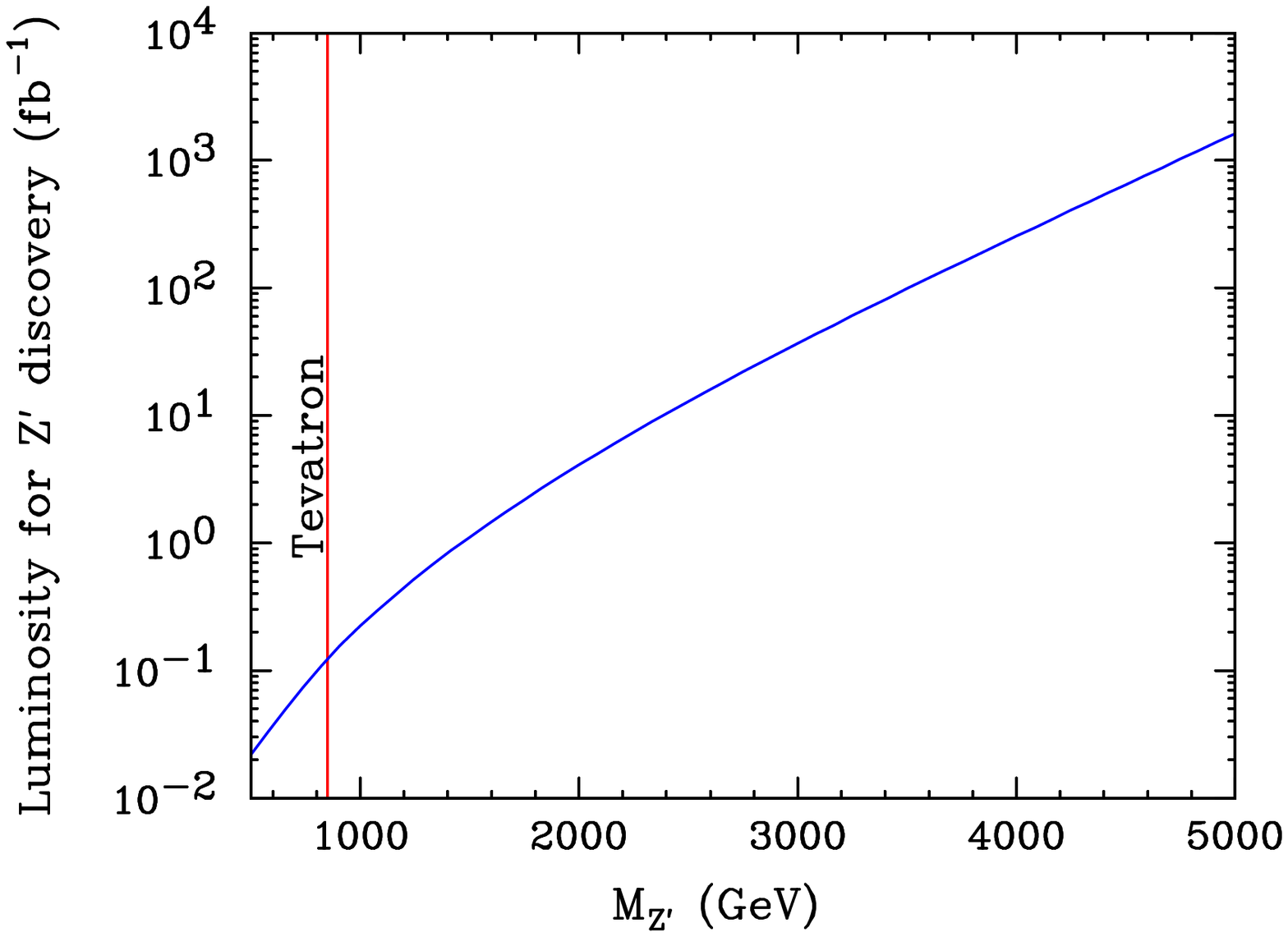}
\end{center}
\caption{(left) Lower bound on $M_{Z'}$ of the DLRM from the Tevatron
dielectron search. (right) Luminosity for $Z'$ discovery by 10 dielectron
events at the LHC.}
\label{fig:collider}
\end{figure}

\newpage
\noindent \underline{\it Odd R particles}~:~ The particles which have odd $R$
are $n$, $h$, $W_R^\pm$, $\phi_R^\pm$, $\Delta_R^\pm$, $\phi_1^\pm$,
$Re(\phi_1^0)$, and $Im(\phi_1^0)$.  Note that $Re(\phi_1^0)$ and
$Im(\phi_1^0)$ are split in mass by the term $Tr(\tilde{\Phi}^\dagger \Delta_L
\Phi \Delta_R^\dagger)$ as the result of $v_5 \neq 0$.  Since $m_\nu$ comes
from $v_5$, this splitting is very small and not enough to qualify either
to be a dark-matter candidate, because its scattering with nuclei through
$Z$ exchange would be much too big for it not to have been detected in
direct-search experiments.  This leaves the lightest $n$ as a viable
dark-matter candidate and we call it a {\it scotino}.  Although it does not
couple to $Z$, it has $Z'$ interactions.  To satisfy the Tevatron search
limits, $M_{Z'}$ should be greater than 850 GeV.  On the other hand,
$\Delta_R$ has no such constraint and the Yukawa interaction $e_R n_R
\Delta_R^+$ may well be the one responsible for the annihilation of $n$ in the
early Universe for which the observed relic abundance of dark matter is
obtained.  Since $\Delta_R$ does not couple to quarks, there is also no
constraint from DM direct-search experiments for these interactions. Note
that the Yukawa interactions $\bar{\nu}_L n_R \phi_1^0$ and $\bar{e}_L n_R
\phi_1^-$ also exist, but are too small because they are proportional to $m_e$.

\newpage
\noindent \underline{\it Scotogenic neutrino mass}~:~ A simple variation of
this model also allows neutrino masses to be radiatively generated by
dark matter, i.e. {\it scotogenic} \cite{m06}.  Instead of $\Delta_L$,
we add a scalar singlet $\chi \sim (1,1,1,0;-1)$, then the trilinear scalar
term $Tr(\Phi \tilde{\Phi}^\dagger) \chi$ is allowed.  Using the soft term
$\chi^2$ to break $L$ to $(-)^L$, a scotogenic neutrino mass is obtained
as shown in Fig.~2.

\begin{figure}[htb]
\begin{center}
\includegraphics[width=0.45\textwidth]{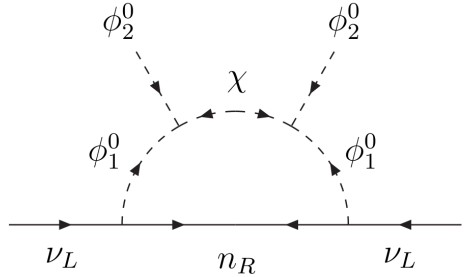}
\end{center}
\caption{One-loop scotogenic neutrino mass.}
\label{fig:neutrinomass}
\end{figure}

\noindent \underline{\it Dark matter}~:~ The lightest $n$ can be a
stable weakly interacting neutral particle.  It is thus a good
candidate for the dark matter of the Universe.  We assume that
$\Delta_R^+$ is much lighter than $W_R^+$ and $Z'$; hence the
dominant annihilation of $n$ is given by $n n \rightarrow e^-_R
e^+_R$ through the exchange of $\Delta^+_R$ with Yukawa coupling
$f_n$.  Using the approximation $\langle \sigma v \rangle \simeq a
+ b v^2$ for the thermally averaged annihilation cross section of
$n$ multiplied by its velocity, we find %
\bea %
a = 0, ~~~ b = \frac{f^4_n}{48 \pi m_n^2} r^2
\left(1-2 r + 2 r^2\right),%
\eea%
where  $r = (1+ w^2)^{-1}$ with $w= m_{\Delta_R^+}/m_n$. Under the
usual assumption that $n$ decoupled from the SM particles in the
early Universe when it became nonrelativistic, its relic
density is given by%
\be %
\Omega_{n} h^2 = \frac{8.76\times 10^{-11}{\rm GeV}^{-2}}{g^{1/2}_{*
(T_F)}(a/x_F+3b/x^2_F)}, ~~~~~ x_F=\ln\frac{0.0955M_{P}
m_{n}(a+6b/x_F)}{\sqrt{g_{*} x_F}},%
\ee%
where $T_F$ is the freeze-out temperature, $x_F=m_{n}/T_F$,
$M_{P}$ the Planck mass, and $g_{*}$ the number of relativistic
degrees of freedom at $T_F$.
%
\begin{figure}[htb]
\begin{center}
\epsfig{file=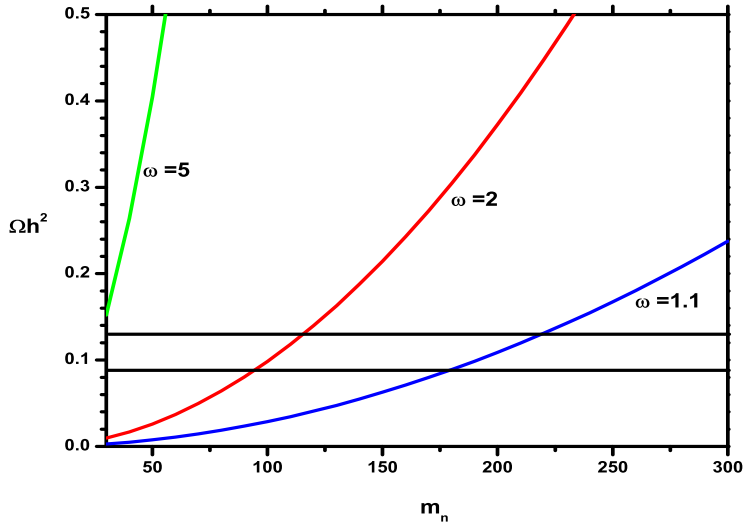, width=11cm, height=8cm, angle=0}
\end{center}
\vskip -1.25cm \caption{Relic density of $n$ as a function of its mass and
$w=m_{\Delta_R^+}/m_n$.  Horizontal lines correspond to the
experimental limits at $1\sigma$.} \label{fig3}
\end{figure}
%
In Fig.~\ref{fig3}, we present the values of the relic abundance
$\Omega_nh^2$ as a function of $m_n$ for the cases $w =
m_{\Delta_R^+}/m_n = 1.1$, 2, and 5, with $f_n = 1$. The measured values 
of $\Omega h^2$ for cold dark matter by the Wilkinson Microwave Anisotropy 
Probe (WMAP) \cite{wmap07}
are obtained for a wide range of $n$ and $\Delta_R^+$ mass values.

Since $n$ always interacts with a lepton in this model, its annihilation
in the Earth's vicinity will produce high-energy electrons and positrons,
which may be an explanation of such recently observed events in the PAMELA
\cite{pamela1} and ATIC \cite{atic} experiments.

\noindent \underline{\it Conclusion}~:~ We have proposed in this paper the
notion that neutrinos and dark-matter fermions (scotinos) exist in parallel
as members of doublets under $SU(2)_L$ and $SU(2)_R$ respectively.  As
such, both interact with leptons: neutrinos through $W_L$ and scotinos
through $W_R$.  The resulting model (DLRM) allows for the definition of a
generalized lepton number and thus $R$ parity in a nonsupersymmetric
context, and has a host of verifiable predictions at the TeV scale.

\newpage
\noindent \underline{\it Acknowledgements}:~
This work was supported in part by the U.~S.~Department of Energy under
Grant No. DE-FG03-94ER40837.  The work of S.K. is supported in part by
ICTP project 30 and the Egyptian Academy of Scientific Research and
Technology.  E.M. thanks The British University in Egypt for its great
hospitality during a recent visit.

\bibliographystyle{unsrt}

\end{document}